\def\ie{{\it i.e.}}

\def\~{{$\tilde{\phantom{a}}$}}

\documentclass [12pt] {article}
\usepackage{epsfig}
\usepackage{color}

\def\thebibliography#1{\section{References}\markboth
 {REFERENCES}{REFERENCES}\list
 {[\arabic{enumi}]}{\settowidth\labelwidth{[#1]}\leftmargin\labelwidth
 \advance\leftmargin\labelsep
 \usecounter{enumi}}
 \def\newblock{\hskip .11em plus .33em minus -.07em}
 \sloppy
 \sfcode`\.=1000\relax}
\def\upcite#1{\raise6pt\hbox{\scriptsize
\cite{#1}}}
  \def\lsim{\mathrel {\vcenter {\baselineskip 0pt \kern 0pt
    \hbox{$<$} \kern 0pt \hbox{$\sim$} }}}
    \def\gsim{\mathrel {\vcenter {\baselineskip 0pt \kern 0pt
    \hbox{$>$} \kern 0pt \hbox{$\sim$} }}}

\def\hline{\noalign{\hrule \vskip2pt}}

\def\|{\ifmmode\Vert\else \char`\|\fi}
\ifx\oldzeta\undefined                          
  \let\oldzeta=\zeta                            
  \def\zzeta{{\raise 2pt\hbox{$\oldzeta$}}}     
  \let\zeta=\zzeta                              
\fi

\ifx\oldchi\undefined                           
  \let\oldchi=\chi                              
  \def\cchi{{\raise 2pt\hbox{$\oldchi$}}}       
  \let\chi=\cchi                                
\fi

\def\frac#1#2{{#1 \over #2}}

\def\half{\ifinner {\scriptstyle {1 \over 2}}
   \else {1 \over 2} \fi}

\def\ave#1{\left\langle#1\right\rangle} 

\def\simge{\mathrel{%
   \rlap{\raise 0.511ex \hbox{$>$}}{\lower 0.511ex \hbox{$\sim$}}}}
\def\simle{\mathrel{
   \rlap{\raise 0.511ex \hbox{$<$}}{\lower 0.511ex \hbox{$\sim$}}}}

\def\buildchar#1#2#3{{\null\!                   
   \mathop#1\limits^{#2}_{#3}                   
   \!\null}}                                    
\def\overcirc#1{\buildchar{#1}{\circ}{}}

\def\slashchar#1{\setbox0=\hbox{$#1$}           
   \dimen0=\wd0                                 
   \setbox1=\hbox{/} \dimen1=\wd1               
   \ifdim\dimen0>\dimen1                        
      \rlap{\hbox to \dimen0{\hfil/\hfil}}      
      #1                                        
   \else                                        
      \rlap{\hbox to \dimen1{\hfil$#1$\hfil}}   
      /                                         
   \fi}                                         %

\def\subrightarrow#1{
  \setbox0=\hbox{
    $\displaystyle\mathop{}
    \limits_{#1}$}
  \dimen0=\wd0
  \advance \dimen0 by .5em
  \mathrel{
    \mathop{\hbox to \dimen0{\rightarrowfill}}
       \limits_{#1}}}                           

\def\overlay#1#2{\ifmmode%
\setbox0=\hbox{$#1$}%
\setbox1=\hbox to\wd0{\hss$#2$\hss}\else%
\setbox0=\hbox{#1}%
\setbox1=\hbox to\wd0{\hss#2\hss}\fi%
#1\hskip-\wd0\box1 }

\def\pmb#1{\leavevmode\setbox0=\hbox{#1}%
\kern-.02em\copy0\kern-\wd0
\kern.04em\copy0\kern-\wd0
\kern-.02em\raise.04em\box0 }

\def\vereq#1#2{\lower3pt\vbox{\baselineskip1.5pt \lineskip1.5pt
\ialign{$\m@th#1\hfill##\hfil$\crcr#2\crcr\sim\crcr}}}

\def\tensor#1{\protect\@ontopof{#1}{\leftrightarrow}{1.15}\mathord{\box2}}
\def\overstar#1{\protect\@ontopof{#1}{\ast}{1.15}\mathord{\box2}}
\def\overdots#1{\protect\@ontopof{#1}{\cdots}{1.0}\mathord{\box2}}
\def\overcirc#1{\protect\@ontopof{#1}{\circ}{1.2}\mathord{\box2}}
\def\loarrow#1{\protect\@ontopof{#1}{\leftarrow}{1.15}\mathord{\box2}}
\def\roarrow#1{\protect\@ontopof{#1}{\rightarrow}{1.15}\mathord{\box2}}

\def\@ontopof#1#2#3{%
{\mathchoice
{\@@ontopof{#1}{#2}{#3}\displaystyle\scriptstyle}%
{\@@ontopof{#1}{#2}{#3}\textstyle\scriptstyle}%
{\@@ontopof{#1}{#2}{#3}\scriptstyle\scriptscriptstyle}%
{\@@ontopof{#1}{#2}{#3}\scriptscriptstyle\scriptscriptstyle}%
}%
}

\def\@@ontopof#1#2#3#4#5{%
\setbox0=\hbox{$#4#1$}%
\setbox1=\hbox{$#5#2$}%
\setbox2=\hbox{}\ht2=\ht0 \dp2=\dp0 %
\ifdim\wd0>\wd1 %
\setbox1=\hbox to\wd0{\hss\box1\hss}%
\mathord{\rlap{\raise#3\ht0\box1}\box0}%
\else   %
\setbox1=\hbox to.9\wd1{\hss\box1\hss}%
\setbox0=\hbox to\wd1{\hss$#4\relax#1$\hss}%
\mathord{\rlap{\copy0}\raise#3\ht0\box1}%
\fi
}%

\def\lambdabar{\protect\@lambdabar}
\def\@lambdabar{%
\relax
\bgroup
\def\@tempa{\hbox{\raise.73\ht0
\hbox to0pt{\kern.25\wd0\vrule width.5\wd0
height.1pt depth.1pt\hss}\box0}}%
\mathchoice{\setbox0\hbox{$\displaystyle\lambda$}\@tempa}%
{\setbox0\hbox{$\textstyle\lambda$}\@tempa}%
{\setbox0\hbox{$\scriptstyle\lambda$}\@tempa}%
{\setbox0\hbox{$\scriptscriptstyle\lambda$}\@tempa}%
\egroup
}

\def\corresponds{{\lower.2ex\hbox{=}}{\rm\kern-.75em^\triangle}}
\def\succsim{\succ\kern-.9em_\sim\kern.3em}
\def\precsim{\prec\kern-1em_\sim\kern.3em}
\def\slantfrac#1#2{\kern1em^{#1}\kern-.3em/\kern-.1em_{#2}}

\begin{document}

\begin{center}
{\Large\bf Hertzian Dipole Radiation via the 
\\
\medskip
Weizs\"acker-Williams Method}
\\

\medskip

Kirk T.~McDonald
\\
{\sl Joseph Henry Laboratories, Princeton University, Princeton, NJ 08544}
\\
Max S.~Zolotorev
\\
{\sl Center for Beam Physics, Lawrence Berkeley National Laboratory,
Berkeley, CA 94720}
\\
(Aug.\ 4, 2003)
\end{center}

\section{Problem}

Use the Weizs\"acker-Williams method to deduce the radiated power, 
and its angular distribution, emitted by an electron of charge $e$ that
undergoes oscillatory motion, $z = z_0 \cos\omega t$.

\section{Solution}

This problem is a continuation of an earlier note on the Weizs\"acker-Williams
method \cite{weizsacker}.

The basic idea of the method is that if a rapidly moving electron is sufficiently
perturbed, it will emit $\alpha = e^2 / \hbar c$ photons per unit bandwidth during
per formation time $t_0$, where the latter is the time it takes for the radiation to
pull one wavelength ahead of the charge.  Thus, the photon number spectrum of the
radiation is
\begin{equation}
{d n \over d\omega d t_0} \approx \alpha.
\label{s0}
\end{equation}

Underlying the method is the fact that the electromagnetic fields of a relativistic
charge are largely transverse to the direction of motion of the charge, and hence
much like a free electromagnetic wave (pulse).

Here, we seek to apply this method to nonrelativistic radiation by an oscillating
charge.  Because the charge has a maximum velocity (relative to the speed of light $c$)
\begin{equation}
\beta = {v_{\rm max} \over c}
= {z_0 \omega \over c}
\label{s1}
\end{equation}
much
less than one, we anticipate that the radiation will not have the full strength
attainable in the relativistic limit.  Rather, the amplitude of the radiation will be
scaled by a factor $\beta$, and so the radiated power will be scaled down
by a factor $\beta^2$.\footnote{A related argument was given in Sec.~IIID of 
\cite{weizsacker} for the case of radiation by an electron passing through a
weak undulator.}

Because the motion of the charge is periodic, the radiation spectrum is a
line at angular frequency $\omega$.  We estimate
that the rate of photon emission at frequency $\omega$ is roughly the same as that in
one unit of bandwidth in the continuum version of the model,

Because the motion of the electron is bounded (with amplitude small compared
to the wavelength $\lambda = 2 \pi c / \omega$), the formation length is
simply a wavelength, and the formation time $t_0$ is simply one period, $ 2\pi / \omega$.

Applying, the basic principle (\ref{s0}) of the Weizs\"acker-Williams method, with the 
modifications described above for nonrelativistic, period motion of the charge,
the number of photons radiated per unit time is
\begin{equation}
{dN \over dt} \approx \alpha \beta^2 {\omega \over 2 \pi}
= {e^2 \over \hbar c} {z_0^2 \omega^2 \over c^2} {\omega \over 2 \pi}\, .
\label{s2}
\end{equation}
The radiated power is $\hbar \omega$ times this:
\begin{equation}
P = \hbar \omega {dN \over dt} \approx {(e z_0)^2 \omega^4 \over 2 \pi  c^3} 
= {p_0^2 \omega^4 \over 2 \pi  c^3}\, . 
\label{s3}
\end{equation}
where $p_0 = e z_0$ is the (maximum) dipole moment of the electron.  The result (\ref{s3})
is about 1/2 that of the usual expression for the time-averaged power from an
oscillation dipole $p = p_0 \cos\omega t$,
\begin{equation}
\ave{P}= {p_0^2 \omega^4 \over 3  c^3}\, . 
\label{s4}
\end{equation}
The factor $2 \pi / 3$ difference between forms (\ref{s3}) and (\ref{s4}) indicates
that the effective bandwidth of the continuum that is ``squeezed'' into the
line spectrum is only about 1/2 unit, rather than one as assumed above.

Some care is required to determine the angular distribution of the radiation from
the Weizs\"acker-Williams perspective.  One might be tempted to argue that the
radiation will be emitted preferentially in the direction of velocity {\bf v} of the
charge, because wavelike fields that surround the charge (its cloud of ``virtual photons'')
have ${\bf E} \times {\bf B}$ largely parallel to {\bf v}.  And indeed, in the
relativistic limit the radiation is emitted primarily in the {\bf v} direction.  
However, the present problem is in the
nonrelativistic limit, in which the average velocity of the charge is zero,
so we must look elsewhere for a description of the angular distribution.

A useful argument is based on the radiation reaction.  Power is emitted by the
oscillating charge, so we expect a back reaction on the charge.  This reaction
must come from the interaction of the charge with the radiation fields.  The
power of the reaction force is
\begin{equation}
P_{\rm react} = {\bf F}_{\rm react} \cdot {\bf v}
= e{\bf E}_{\rm react} \cdot {\bf v}
= e{\bf E}_{\rm rad} \cdot {\bf v}.
\label{s5}
\end{equation}
To provide the required radiation reaction, the electric field ${\bf E}_{\rm rad}$
of the radiation
must have a large component along the direction of {\bf v}, \ie, along the $z$
axis in the present problem.  This indicates that the electric field of
radiation emitted at angle $\theta$ measured with respect to
the $z$ axis varies as $\sin\theta$ (rather than $\cos\theta$ as would be too
naively inferred from the previous paragraph).

The radiated power goes as the square of the electric field, so the angular
distribution corresponding to eq.~(\ref{s3}) is
\begin{equation}
{d P \over d \Omega} = {3 p_0^2 \omega^4 \over 8 \pi^2  c^3} \sin^2\theta. 
\label{s6}
\end{equation}
Or, applying the factor $2 \pi / 3$, we recover the usual Hertzian expression,
\begin{equation}
{d \ave{P} \over d \Omega} = {p_0^2 \omega^4 \over 4 \pi  c^3} \sin^2\theta. 
\label{s7}
\end{equation}

\bigskip

K.T.M.\ wishes to thank Dan Handelsman, N2DT, and David Jeffries for the e-discussions
that encouraged a different way of thinking about radiation from antennas.

\end{document}